\documentclass[
    ,final            
  ]
  {aipproc}

\layoutstyle{6x9}

\begin{document}

\title[Young stellar objects]{Young stellar objects from soft to hard X-rays}


\author{Manuel G\"udel}{
  address={Institute of Astronomy, ETH Zurich, 8093 Zurich, Switzerland}
}

\begin{abstract}
 Magnetically active stars are the sites of efficient particle acceleration and plasma heating, 
 processes that have been studied in detail in the solar corona. Investigation of such processes 
 in young stellar objects is much more challenging due to various
 absorption processes. There is, however, evidence for violent magnetic energy release in 
 very young stellar objects. The impact on young stellar environments (e.g., circumstellar
 disk heating and ionization, operation of chemical networks, photoevaporation) may be substantial.
 Hard X-ray devices like those carried on Simbol-X will establish a basis for detailed studies
 of these processes.
\end{abstract}

\maketitle


\section{Introduction}

X-ray radiation from young stars are fundamentally important for  molecular stellar environments. X-rays ionize 
protostellar envelopes and circumstellar 
disks  \citep{glassgold97}. In the presence of weak magnetic fields, ionized disk surface layers 
grow unstable to the magnetorotational instability \citep{balbus91}, currently accepted 
as the most promising driver of  accretion through disks. Further, X-ray irradiation of circumstellar 
molecular gas drives important chemistry, including production or destruction of water \citep{glassgold04, 
stauber06}. X-rays heat disk surfaces to temperatures of several thousand K \citep{glassgold04}, 
giving rise to (partial) photoevaporation of, and therefore mass loss from, the innermost disk regions
\citep{ercolano08}. 

Stellar X-ray astronomy has concentrated on the {\it soft} X-ray range (SXR, $\approx$~0.1--10~keV), 
where {\it thermal} bremsstrahlung and line radiation dominate. Harder X-rays $>$10~keV (HXR)
are somewhat of a {\it terra incognita} in stellar astronomy, notwithstanding a few firm detections 
beyond $\approx 20$~keV and marginal claims for detections of non-thermal HXR.
Simbol-X promises a breakthrough for stellar HXR, thus opening a window to deeply embedded, young
X-ray sources and to the physics of magnetic energy release in young, magnetically active stars 
\citep{argiroffi08, maggio08, micela08a, micela08b, sciortino08b}.

\section{Magnetic Fields in Protostars?}

It is essentially unknown when during star formation {\it stellar} magnetic fields 
first appear, and whether they are fossil or are generated by internal dynamos.
Our understanding of  magnetic-field induced high-energy environments of the youngest
stars is very limited.

A rather comprehensive picture of the X-ray characteristics of T Tauri stars (TTS) is available,
especially from large recent surveys with XMM-Newton and Chandra \citep{getman05, guedel07a, sciortino08a}. 
In short, TTS X-ray emission mostly originates from magnetic coronae, with characteristics similar 
to more evolved active main-sequence stars. Their X-rays saturate at a level of  
log($L_{\rm X}/L_{\rm bol}) \approx -3.5$. Because for a typical pre-main sequence association
$L_{\rm bol}$ roughly correlates with stellar mass $M_*$, one also finds a distinct correlation
between $L_{\rm X}$ and $M_*$, $\log L_{\rm X} \propto M_*^{1.7\pm 0.1}$ \citep{telleschi07}. Flaring is
common in TTS \citep{stelzer07}, the most energetic examples reaching temperatures of 
$\approx 10^8$~K \citep{imanishi01}.

The complex stellar environments of protostars and TTS may lead to further X-ray sources.
Accretion streams may form shocks at the stellar surface, 
producing very soft X-rays. The evidence is twofold: unusually high electron densities
have been inferred from spectral line ratios from a handful of TTS \citep{kastner02},
and a general ``soft excess'', defined by an anomalously high ratio of O\,{\sc vii}~$\lambda 21.6$
to O\,{\sc viii}~$\lambda 19.0$ line fluxes \citep{guedel07c}. 
Jets and outflows may shock heat some gas to $>$1~MK, both in
Herbig-Haro objects far away from the driving stars \citep{pravdo01} and internal to jets
very close to the stars \citep{guedel07b}.
These  X-ray sources are also soft, typically detected 
only below $\approx$1-2~keV. This is not surprising given the energy available from shock velocities of 
no more than a few hundred km~s$^{-1}$. 

The picture is much less complete for earlier phases of star formation owing to
strong X-ray attenuation. Non-thermal radio emission from
Class I protostars suggests the presence not only of magnetic fields but also accelerated
particles, both in distant regions of outflows (e.g., \citep{ray97}) and within
the stellar  corona \citep{smith03}.

For  ``Class 0 objects'' (the earliest phase of a forming star), a few promising 
candidates but no definitive cases have been 
reported (e.g., \citep{hamaguchi05}). \citet{giardino07} summarize several Class 0 non-detections,
with a ``stacked Class 0 data set'' corresponding to 540~ks of Chandra ACIS-I 
exposure time still giving no indication for a detection. In the absence of detailed information on
the absorbing gas column densities or the intrinsic spectral properties, an interpretation within
an evolutionary scenario is difficult.

The situation is more favorable for more evolved Class I protostars although sample statistics are biased by 
strong X-ray attenuation, favoring detection of the most luminous and the hardest sources. A 
comprehensive study is available for the Orion region \citep{prisinzano08}.
The X-ray luminosities increase
from  protostars to TTS by about an order of magnitude, although the situation is unknown below
1-2~keV. No significant trend is found for the electron temperatures, which are similarly
high ($\approx$1--3~keV) for all detected classes.

Protostellar HXR detections with Simbol-X will be particularly interesting:
emission above $\approx 5-10$~keV is mostly unabsorbed, permitting unbiased 
studies of deeply embedded protostars, thus pinpointing the earliest appearance of magnetic activity
and high-energy stellar environments. Such observations will also be crucial to understand 
magnetic energy release in these youngest objects, as discussed below.

\section{Flare Physics in Young Stars}

\begin{figure}\label{fig1}
  \hbox{
  \includegraphics[height=.25\textheight]{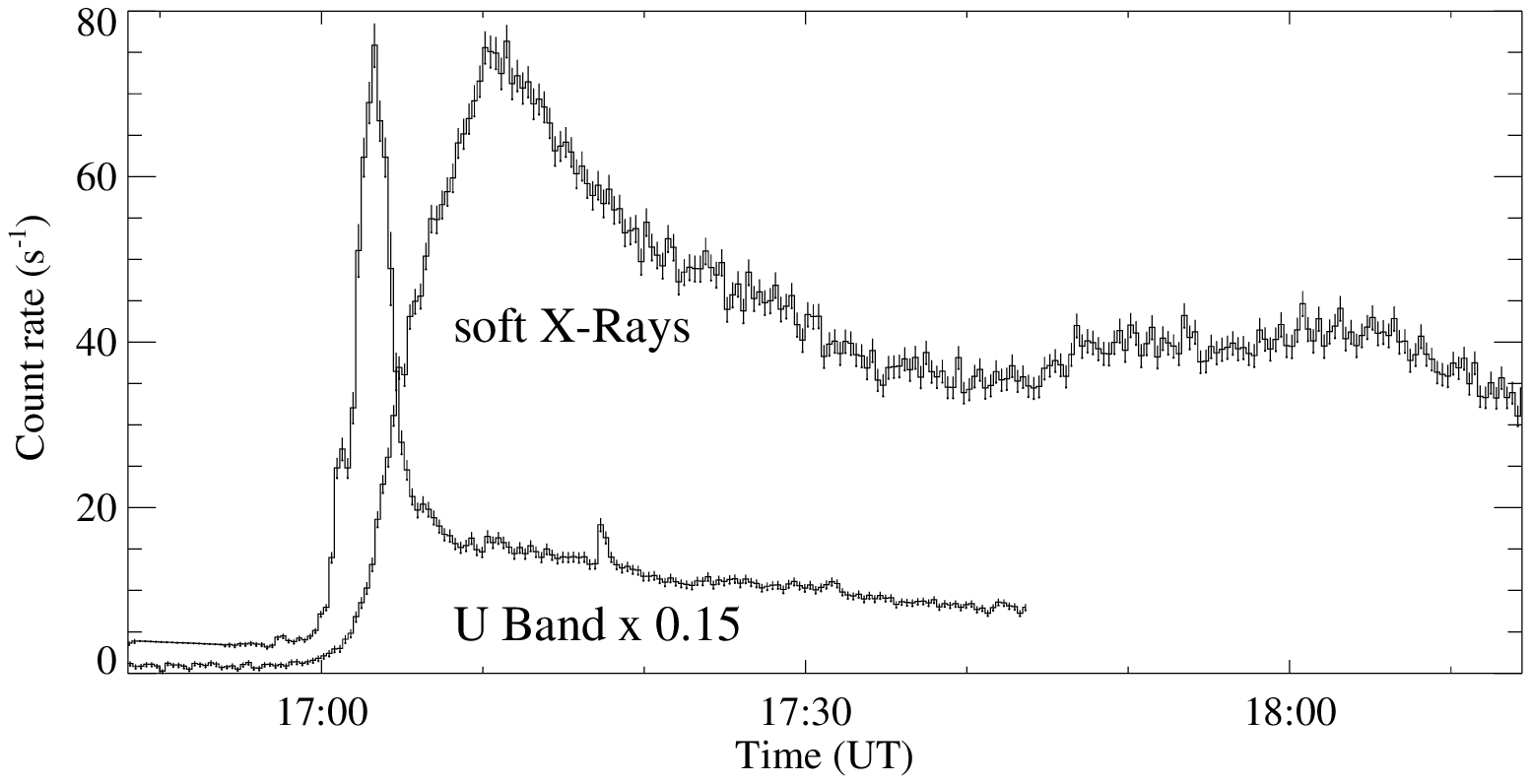}
  \includegraphics[height=.25\textheight]{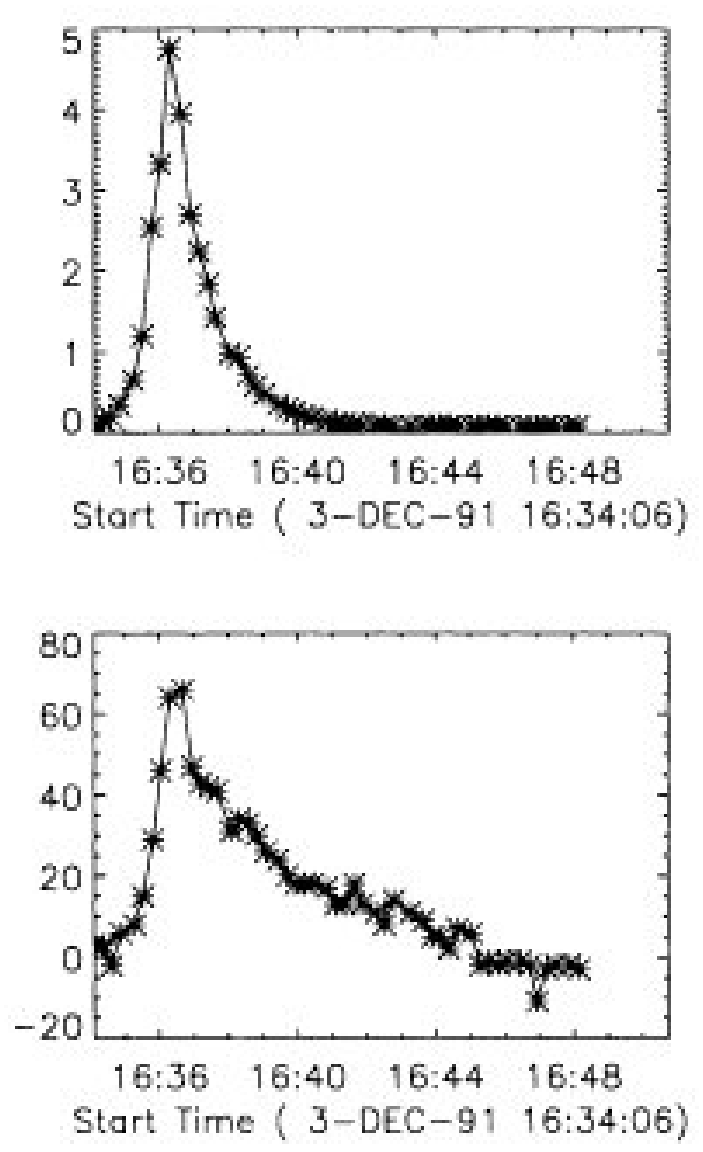}
  }
  \caption{Left (a): Large stellar flare on Proxima Cen, showing an initial U-band burst defining the impulsive 
  phase, and a delayed soft X-ray flare defining the gradual phase \citep{guedel02b}. - 
  Right (b): Correlation between a solar white light (upper) and a HXR burst (lower panel; from \citep{hudson92}).}
\end{figure}

Magnetic energy release and the ensuing electromagnetic spectrum have been studied in detail for
the solar corona \citep{dennis85, lin02}. The standard flare scenario posits that non-potential
coronal magnetic fields reconnect and thus release kinetic and thermal energy. While traveling along closed magnetic fields and
colliding in denser chromospheric layers, accelerated particles emit prompt gyrosynchrotron, optical, and ultraviolet radiation 
and non-thermal hard X-rays. 
The energy deposition explosively heats gas to $>$10~MK, and the resulting overpressure drives 
the newly heated gas along the magnetic field lines into the corona. Soft X-rays are emitted by the hot plasma
accumulating in coronal loops, with a characteristic delay with respect to the prompt non-thermal emission (Fig.~\ref{fig1}a).

HXR emission provides crucial diagnostics for the particle acceleration process in the solar corona and therefore
the primary energy release mechanisms (Fig.~\ref{fig1}b), yet very little is known about HXR emission from active stars. Detections of a few
sources up to several tens of keV with BeppoSAX  \citep{favata99, pallavicini00, franciosini01} turn out to be compatible 
with the extended thermal spectrum also
seen in the SXR range. Perhaps the most promising case so far has been reported from Swift observations of the active 
binary II Peg \citep{osten07}. During a giant flare, the observed X-ray spectrum up to at least 100~keV is compatible with a power-law
electron distribution although a thermal, very hot Maxwellian distribution cannot be excluded. 
 
There is ample complementary evidence for non-thermal electron populations in stellar coronae, including young
stellar objects. Radio gyrosynchrotron emission has been detected in many active stars during flares
and quiescent periods; the quiescent radio emission correlates with the quiescent SXR radiation, suggesting a causal 
relation between particle acceleration and heating in the same way as in solar flares
\citep{guedel02a}. This would suggest the presence of continuous HXR emission as well, as discussed 
below.

High-energy electrons are also capable of ejecting inner-shell electrons in cool gas, producing the analog of
fluorescent emission induced by photon irradiation. Although the 6.4~keV line of ``cold iron'' occasionally detected in 
young stellar objects may indeed be from fluorescence in X-ray irradiated circumstellar disks, new
evidence suggests a role for electron impact, both in giant flares in evolved stars
\citep{osten07} and in a protostellar flare  in which the 6.4~keV line peaked during the rise phase of the soft X-rays
\citep{czesla07}. 

\begin{figure}
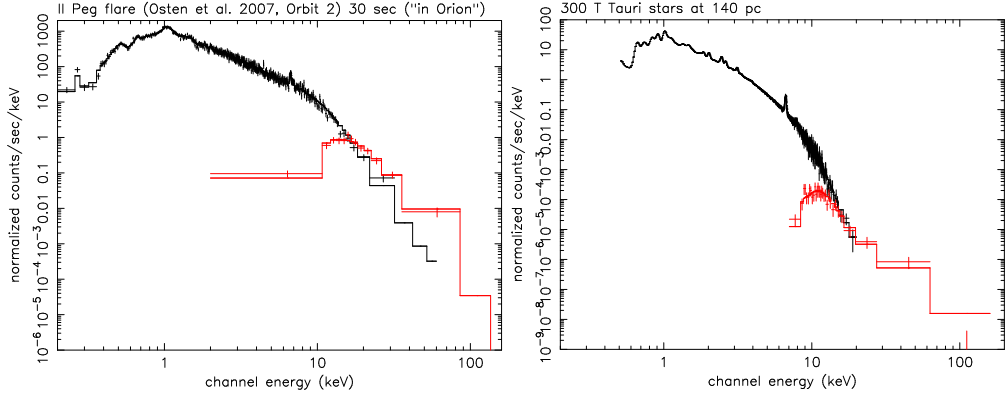
\label{fig2}
  \hbox{
  \includegraphics[height=.3\textheight,angle=-90]{m_guedel1_f2a.ps}
  \includegraphics[height=.3\textheight,angle=-90]{m_guedel1_f2b.ps}
  }
  \caption{Simbol-X simulations. Left (a): II Peg flare \citep{osten07} at the distance of the Orion Nebula (3~ks of 
  exposure time).
  Right (b): Summed 200~ks spectrum of 300 typical TTS in  Taurus, assuming continuous flaring. }
\end{figure}

Simbol-X will clarify these issues in detail. Fig.~\ref{fig2}a shows a spectral simulation of the II Peg flare \citep{osten07}
as if it had occurred at the distance of the Orion Nebula, exposed for 3~ks. The power-law tail suggested by \citet{osten07} 
is included in the model and dominates all emission above $\approx$30~keV. Although such giant flares are exceedingly rare on 
an individual star, a cluster with $>1000$ young stars such as the Orion Nebula Cluster may offer a realistic chance for Simbol-X
to study several such events.

\section{Hard X-rays from continuously flaring coronae?}

Solar observations show evidence for
small-scale flare events occurring in the  corona at any time (e.g., \citep{lin84, krucker98}). 
Their distribution in energy follows a  power law,
\begin{equation}\label{e:powerlaw}
\frac{dN}{dE} = k E^{-\alpha} 
\end{equation}
where $dN$ is the number of flares per unit time with a total 
energy (thermal or radiated) in the  interval [$E,E+dE$]. If the power-law index $\alpha$ is
$\ge 2$, then the energy integration 
\begin{equation}\label{e:powerlaw2}
P_{\rm tot} = \int_{E_{\rm min}}^{E_{\rm max}}{dN\over dE}EdE  \approx 
    {k\over \alpha-2}E_{\rm min}^{-(\alpha - 2)}
\end{equation}    
(assuming $E_{\rm min} \ll E_{\rm max}$ and $\alpha > 2$ for the last 
approximation) diverges for $E_{\rm min} 
\rightarrow 0$, i.e., by extrapolating the power law 
to sufficiently small flare energies, {\it any} energy release
power can be attained \citep{hudson91}. Solar studies have repeatedly
resulted in $\alpha$ values on the order of 1.6--1.8 for ordinary
solar flares \citep*{crosby93} although some statistical investigations
suggest $\alpha = 2.0 - 2.6$ for small flares in the quiet solar corona 
\citep{krucker98,parnell00}. Stellar studies in the EUV or soft X-ray range have
predominantly converged to $\alpha$ between 2 and 3, and this is specifically 
true for large samples of pre-main sequence stars (e.g., \citep{guedel03, stelzer07}).

An ultimate test of this model would be the detection of continuous hard, non-thermal X-ray 
emission that should accompany the energy release diagnosed so far in the soft X-ray range (e.g., 
\citep{micela08a}). I propose
in the following a method to estimate the non-thermal hard X-ray production based on some basic relations
derived from the Sun and magnetically active stars.
For a complementary discussion and application, I refer to \citet{micela08a} and \citet{caramazza09}.

The constant $k$ in Eqns.~\eqref{e:powerlaw} and \eqref{e:powerlaw2} can be found from observations 
of the flare occurrence rate in young, magnetically active stars.  
The occurrence rate $N(>10^{32})$ of flares exceeding a radiative soft X-ray output of $10^{32}$~erg is,
for a statistical sample,
\begin{equation}
N(>10^{32}) \approx 6.9\times 10^{-34} L_{\rm X}\quad [{\rm s}^{-1}]
\end{equation}
\citep{audard00} which agrees within a factor of $\approx 2-3$ with the observations of $N(>10^{32})$ 
for very active young solar analogs if $\alpha \approx 2-2.4$ (as derived from the flare-energy 
distributions). Note that  $N(>10^{32})$ is proportional to the average soft X-ray luminosity $L_{\rm X}$ of the star.

{\it Solar} soft X-ray flares are accompanied by non-thermal hard X-rays. The respective peak fluxes 
$F_{\rm s}$ and $F_{\rm h}$ are correlated albeit in a non-linear way \citep{battaglia05, isola07}:
\begin{equation}
F_{\rm h} = {\rm const}F_{\rm s}^{\sigma}
\end{equation}
with $\sigma = 1.2$ \citep{battaglia05} or $\sigma = 1.37$ \citep{isola07}. As  $F_{\rm s}$ refers to
the 1.6-12.4~keV channel of the Geostationary Operational Environmental Satellites (GOES), a conversion
to the broader soft X-ray range of, say, 0.3-10~keV, is needed; for typical spectra of classical
TTS, the GOES flux is about $z = 0.29$ times the total soft X-ray flux.
Converting the non-linear relation for measured fluxes at Earth to luminosities of the solar corona, one 
has\begin{equation}
L_{\rm h} = fz^{\sigma}L_{\rm s}^{\sigma}
\end{equation}
where $f = 2.3\times 10^{-11}$ (cgs) from the \citet{battaglia05} study. The effective observation of (quasi-)
continuous SXR or HXR emission records the total radiated energy $E_{\rm s}$,  $E_{\rm h}$  integrated
in time from each flare rather than their peak luminosities. Introducing a characteristic decay time 
$\tau_{\rm s}$ and $\tau_{\rm h}$ for the soft and hard flare light curve so that 
$E_{\rm s,h} = \tau_{\rm s,h}L_{\rm s,h}$, one finds for the {\it average} power in hard X-rays
\begin{eqnarray}
L_{\rm H}  &=& \int_{\rm E_{h,0}}^{\infty} {dN\over dE_{\rm h}}E_{\rm h}dE_{\rm h} \\
           &=& {k\over \alpha - \sigma - 1}\tau_{\rm s}^{-(\alpha-1)}
	   		                   \tau_{\rm h}^{(\alpha-1)/\sigma}
	   		                   f^{(\alpha-1)/\sigma}
	   		                   z^{\alpha-1}
			                   E_{\rm h,0}^{-(\alpha-\sigma-1)/\sigma}
\end{eqnarray}
where $E_{\rm h,0}$ is the lowest flare energy required. It can be found by noting 
that the corresponding $L_{\rm h,0} = fz^{\sigma}L_{\rm s,0}^{\sigma}$ and that the integration
of the {\it soft} energies should (according to the flare hypothesis) equal the average soft X-ray
luminosity, $L_{\rm X}$, from which
\begin{equation}
E_{\rm s,0} = \left[ {(\alpha -2)L_{\rm X}\over k}\right]^{1/(-\alpha+2)}.
\end{equation}
After some further manipulations, one finds
\begin{equation}
L_{\rm H} = \left(6.9\times 10^{-34}10^{32(\alpha-1)} {\alpha-1\over \alpha-2} \right)^{(1-\sigma)/(2-\alpha)}
            {\alpha -2\over \alpha-\sigma-1} {\tau_{\rm h}\over \tau_{\rm s}^{\sigma}} fz^{\sigma} L_{\rm X}.
\end{equation}
Using parameters from \citet{battaglia05} ($\sigma = 1.2, f = 2.3\times 10^{-11}$), $z = 0.29$, $\alpha = 2.4$
and reasonable decay times of $\tau_{\rm s} = 3000$~s and $\tau_{\rm h} = 1000$~s (see \citep{guedel03} and
\citep{isola07}), one finds $L_{\rm H} = 8.5\times 10^{-7}L_{\rm X}$. Note that in the work by 
\citet{battaglia05}, the hard X-ray flux is ``per keV'' at 35~keV. 
The total hard emission above $\approx 20$~keV could reasonably be a factor of perhaps 50 higher, 
i.e. $L_{\rm H,  tot} \approx 4\times 10^{-5}L_{\rm X}.$

Note that $L_{\rm H} \propto L_{\rm X}$ despite the non-linearities between flare peak powers in SXR and HXR, 
although the (different) flare decay times  become important. The expected radiation 
may be easily detectable with Simbol-X from young stellar clusters if a sufficient
number of stars (each assumed to produce equal emission) contribute. Fig.~\ref{fig2}b shows a 200~ks simulation for
Simbol-X assuming 300 stars with characteristics of an average TTS in the Taurus Molecular Cloud
($L_{\rm X} = 1.2\times 10^{30}$~erg~s$^{-1}$, $\alpha = 2.4$) at a distance of 140~pc. Hard radiation can be 
detected up to 100~keV. The many uncertainties entering the above derivation of course make this an order-of-magnitude 
estimate only.





\IfFileExists{\jobname.bbl}{}
 {\typeout{}
  \typeout{******************************************}
  \typeout{** Please run "bibtex \jobname" to optain}
  \typeout{** the bibliography and then re-run LaTeX}
  \typeout{** twice to fix the references!}
  \typeout{******************************************}
  \typeout{}
 }


\begin{thebibliography}{99}

\bibitem[Argiroffi et al.(2008)]{argiroffi08}C.~Argiroffi, G.~Micela, and A.~Maggio, \emph{Mem.S.A.It.} \textbf{79}, 219 (2008).
\bibitem[Audard et al.(2000)]{audard00}M.~Audard, M.~G\"udel, J.~J.~Drake, and V.~L.~Kashyap,  \emph{ApJ}  \textbf{541}, 396 (2000).
\bibitem[Balbus \& Hawley(1991)]{balbus91}S.~A.~Balbus, and J.~F.~Hawley,  \emph{ApJ} \textbf{376}, 214 (1991).  
\bibitem[Battaglia et al.(2005)]{battaglia05}M.~Battaglia, P.~C.~Grigis, and A.~O.~Benz, 2005, \emph{A\&A} \textbf{439}, 737 (2005).
\bibitem[Caramazza et al.(2009)]{caramazza09}M.~Caramazza, J.~J.~Drake, G.~Micela, and E.~Flaccomio, these proceedings (2009).
\bibitem[Crosby et al.(1993)]{crosby93}N.~B.~Crosby, M.~J.~Aschwanden, and B.~R.~Dennis, \emph{Solar Phys.} \textbf{143}, 275 (1993).
\bibitem[Czesla \& Schmitt(2007)]{czesla07}S.~Czesla, and J.~H.~M.~M.~Schmitt, \emph{A\&A} \textbf{470}, L13 (2007).
\bibitem[Dennis(1985)]{dennis85}B.~R.~Dennis,  \emph{Solar Phys.} \textbf{100}, 465 (1985).
\bibitem[Ercolano et al.(2008)]{ercolano08}B.~Ercolano, J.~J.~Drake, J.~C.~Raymond, and C.~C.~Clarke,  \emph{ApJ} \textbf{688}, 398 (2008).
\bibitem[Favata et al.(1999)]{favata99}F.~Favata, and J.~H.~M.~M.~Schmitt, \emph{A\&A} \textbf{350}, 900 (1999).
\bibitem[Franciosini et al.(2001)]{franciosini01}E.~Franciosini, R.~Pallavicini, and G.~Tagliaferri,  \emph{A\&A} \textbf{375}, 196 (2001).
\bibitem[Getman et al.(2005)]{getman05}K.~V.~Getman, E.~Flaccomio, P.~S.~Broos, et al.,  \emph{ApJS} \textbf{160}, 319 (2005) .
\bibitem[Giardino et al.(2007)]{giardino07}G.~Giardino, F.~Favata, G.~Micela, et al., \emph{A\&A} \textbf{463}, 275 (2007).
\bibitem[Glassgold et al.(1997)]{glassgold97}A.~E.~Glassgold, J.~Najita, and J.~Igea,  \emph{ApJ} \textbf{480}, 344 (1997).
\bibitem[Glassgold et al.(2004)]{glassgold04}A.~E.~Glassgold, J.~Najita, and J.~Igea,  \emph{ApJ} \textbf{615}, 972 (2004). 
\bibitem[G\"udel(2002)]{guedel02a}M.~G\"udel,  \emph{ARA\&A} \textbf{40}, 217 (2002).  
\bibitem[G\"udel \& Telleschi(2007)]{guedel07c}M.~G\"udel, and A.~Telleschi,  \emph{A\&A} \textbf{474}, L25 (2007). 
\bibitem[G\"udel et al.(2002)]{guedel02b}M.~G\"udel, M.~Audard, S.~L.~Skinner, and M.~I.~Horvath, \emph{ApJ} \textbf{580}, L73 (2002).
\bibitem[G\"udel et al.(2003)]{guedel03}M.~G\"udel, M. Audard, V.~L.~Kashyap, et al., \emph{ApJ} \textbf{582}, 423 (2003).
\bibitem[G\"udel et al.(2007a)]{guedel07a}M.~G\"udel, K.~R.~Briggs, K.~Arzner, et al.,  \emph{A\&A} \textbf{468}, 353 (2007a).
\bibitem[G\"udel et al.(2007b)]{guedel07b}M.~G\"udel, A.~Telleschi, M.~Audard, et al.,  \emph{A\&A} \textbf{468}, 515 (2007b).
\bibitem[Hamaguchi et al.(2005)]{hamaguchi05}K.~Hamaguchi, M.~F.~Corcoran, R.~Petre, et al., \emph{ApJ} \textbf{623}, 291 (2005).
\bibitem[Hudson(1991)]{hudson91}H.~S.~Hudson, \emph{Solar Phys.} \textbf{133}, 357 (1991).
\bibitem[Hudson et al.(1992)]{hudson92}H.~S.~Hudson, L.~W.~Acton, T.~Hirayama, and Y.~Uchida, \emph{PASJ} \textbf{44}, L77 (1992).
\bibitem[Imanishi et al.(2001)]{imanishi01}K.~Imanishi, K.~Koyama, and Y.~Tsuboi, \emph{ApJ} \textbf{557}, 747 (2001).
\bibitem[Isola et al.(2007)]{isola07}C.~Isola, F.~Favata, G.~Micela, and H.~S.~Hudson, \emph{A\&A} \textbf{472}, 261 (2007).
\bibitem[Kastner et al.(2002)]{kastner02}J.~H.~Kastner, D.~P.~Huenemoerder, N.~S.~Schulz, et al., \emph{ApJ} \textbf{567}, 434 (2002).
\bibitem[Krucker \& Benz(1998)]{krucker98}S.~Krucker, and A.~O.~Benz, \emph{ApJ} \textbf{501}, L213 (1998).
\bibitem[Lin et al.(1984)]{lin84}R.~P.~Lin, R.~A.~Schwartz, S.~R.~Kane, et al., \emph{ApJ} \textbf{283}, 421 (1984).
\bibitem[Lin et al.(2002)]{lin02}R.~P.~Lin, B.~R.~Dennis, G.~J.~Hurford, et al., \emph{Solar Phys.} \textbf{210}, 3 (2002).
\bibitem[Maggio(2008)]{maggio08}A.~Maggio,  \emph{Mem.S.A.It.} \textbf{79}, 186 (2008).
\bibitem[Micela \& Caramazza(2008)]{micela08a}G.~Micela, and M.~Caramazza, \emph{Mem.S.A.It.} \textbf{79}, 266 (2008).
\bibitem[Micela et al.(2008)]{micela08b}G.~Micela, F.~Favata, G.~Giardino, and S.~Sciortino, \emph{Mem.S.A.It.} \textbf{79}, 264 (2008).
\bibitem[Osten et al.(2007)]{osten07}R.~A.~Osten, S.~Drake, J.~Tueller, et al., \emph{ApJ} \textbf{654}, 1052 (2007).
\bibitem[Pallavicini et al.(2000)]{pallavicini00}R.~Pallavicini, G.~Tagliaferri, and A.~Maggio, \emph{Adv. Space Res.}, \textbf{25}, 517 (2000).
\bibitem[Parnell \& Jupp(2000)]{parnell00}C.~E.~Parnell, and P.~E.~Jupp, \emph{ApJ} \textbf{529}, 554 (2000).
\bibitem[Pravdo et al.(2001)]{pravdo01}S.~H.~Pravdo, E.~D.~Feigelson, G.~Garmire, et al.,  \emph{Nature}, \textbf{413}, 708  (2001).
\bibitem[Prisinzano et al.(2008)]{prisinzano08}L.~Prisinzano, G.~Micela, E.~Flaccomio, et al., \emph{ApJ} \textbf{677}, 401 (2008).
\bibitem[Ray et al.(1997)]{ray97}T.~P.~Ray, T.~W.~B.~Muxlow, D.~J.~Axon, et al., \emph{Nature}, \textbf{385}, 415 (1997).
\bibitem[Sciortino et al.(2008a)]{sciortino08a}S.~Sciortino, \emph{Astron. Nachrichten} \textbf{329}, 214 (2008a).
\bibitem[Sciortino et al.(2008b)]{sciortino08b}S.~Sciortino, \emph{Mem.S.A.It.} \textbf{79}, 192 (2008b).
\bibitem[Smith et al.(2003)]{smith03}K.~Smith, M.~Pestalozzi, M.~G\"udel, et al., \emph{A\&A} \textbf{406}, 957 (2003).
\bibitem[St\"auber et al.(2006)]{stauber06}P.~St\"auber, J.~K.~Jorgensen, E.~F.~van Dishoeck, et al., \emph{A\&A} \textbf{453}, 555 (2006).
\bibitem[Stelzer et al.(2007)]{stelzer07}B.~Stelzer, E.~Flaccomio, K.~Briggs, et al., \emph{A\&A} \textbf{468}, 463 (2007).
\bibitem[Telleschi et al.(2007)]{telleschi07}A.~Telleschi, M.~G\"udel,  K.~R.~Briggs, et al., \emph{A\&A} \textbf{468}, 425 (2007).

\end{thebibliography}
\end{document}